\documentclass[10pt,twocolumn,letterpaper]{article}

\usepackage[letterpaper,top=0.72in,bottom=0.78in,left=0.67in,right=0.67in,
  columnsep=0.24in]{geometry}
\usepackage[T1]{fontenc}
\usepackage[utf8]{inputenc}
\usepackage{amsmath}
\usepackage{newtxtext,newtxmath}
\usepackage[hyphens]{url}
\usepackage{graphicx}
\usepackage{booktabs}
\usepackage{caption}
\usepackage[numbers,sort&compress]{natbib}
\usepackage{microtype}
\usepackage{authblk}
\usepackage{xurl}
\usepackage[colorlinks=true,allcolors=blue]{hyperref}

\hypersetup{
  pdftitle={Defusing the Trigger: Tail-Risk-Informed Attention Rebalancing for LLM Backdoor Mitigation},
  pdfauthor={Kaisheng Fan, Yishu Gao, Xunzhu Tang, Tegawend\'e F. Bissyand\'e, Weizhe Zhang}
}

\title{Defusing the Trigger: Tail-Risk-Informed Attention Rebalancing\\ for LLM Backdoor Mitigation}

\author[1]{Kaisheng Fan}
\author[1]{Yishu Gao}
\author[3]{Xunzhu Tang}
\author[3]{Tegawend\'e F. Bissyand\'e}
\author[1,2]{Weizhe Zhang\thanks{Corresponding author.}}
\affil[1]{School of Cyber Science and Technology, Harbin Institute of Technology, Harbin, China\\
\texttt{fankaisheng@stu.hit.edu.cn, gaoyishu@stu.hit.edu.cn, wzzhang@hit.edu.cn}}
\affil[2]{Department of New Networks, Peng Cheng Laboratory, Shenzhen, China}
\affil[3]{SnT, University of Luxembourg, Luxembourg City, Luxembourg\\
\texttt{tegawende.bissyande@uni.lu, xunzhu.tang@uni.lu}}
\date{}

\begin{document}

\maketitle
\raggedbottom

\begin{abstract}
Backdoored large language models (LLMs) can exhibit attacker-specified behavior at inference time while retaining normal performance on benign inputs. Existing mitigation methods often require parameter updates and trusted clean data, or rely on auxiliary generation and repeated model execution, complicating their deployment. Across diverse backdoor mechanisms, we observe that successful activations tend to exhibit stronger tail concentration in attention over semantic-content tokens than benign inputs and unsuccessful trigger activations. This pattern provides a sample-internal control signal for selectively regulating suspicious attention dynamics. We propose TIARA, a tail-risk-informed attention rebalancing approach for inference-time LLM backdoor mitigation. TIARA filters structural attention sinks, aggregates sparse high-concentration events across attention rows and heads, and converts the resulting risk signal into selective content-domain power smoothing and adaptive attention-mass reallocation. A constrained reconstruction then writes valid attention distributions back before value aggregation. TIARA requires no parameter updates, auxiliary generation, an additional complete target-model pass, or a deployment-time clean reference set. 
We evaluate TIARA across four distinct backdoor paradigms on dense, reasoning-oriented, and sparse mixture-of-experts LLMs.
Across the three model families, TIARA reduces the average macro ASR to 11.5\%, outperforming the strongest no-update inference-time baseline by 7.2 percentage points while limiting clean-task metric degradation to at most 3.8 percentage points. Under standardized profiling, TIARA adds 12.9\% end-to-end latency over a matched eager-attention baseline; the current unfused path is 23.3\% slower than fused SDPA.
Overall, TIARA establishes sample-conditional attention rebalancing as a practical inference-time control layer for mitigating attention-concentrated LLM backdoors.
\end{abstract}

\section{Introduction}
\label{sec:introduction}

Large language models (LLMs) are increasingly deployed as reusable
components whose operators may not control the original training or
adaptation pipeline. Backdoors implanted through data poisoning,
instruction tuning, parameter editing, or reasoning-path manipulation can
induce attacker-specified behavior while largely preserving benign
performance
\cite{gu2017badnets,kurita2020weight,li2024badedit,
yan2024backdooring,xiang2024badchain,yan2025embedx}. Once a compromised
model reaches deployment, complete retraining or replacement may be
impractical, making inference-time mitigation a critical last line of
defense.

Existing defenses often impose substantial deployment costs. Offline
purification requires trusted data, parameter updates, and
checkpoint-specific optimization, while no-update methods filter inputs,
prepend demonstrations, rewrite prompts, or execute the target model
repeatedly
\cite{liu2018fine,min2024crow,zeng2024beear,zhao2025unlearning,
qi2021onion,liu2024causality,mo2025test,ouyang2025llmbd}.
Attention-based defenses provide a closer interface: PURE prunes and
normalizes suspicious heads, HeadAlign detects cross-head anomalies and
fine-tunes the model, and X-GRAAD combines attention with input gradients
to localize trigger tokens
\cite{zhao2024defense,jin2026anomalous,das2025unmasking}.
Backdoor Attribution uses offline causal analysis to locate sparse
backdoor-relevant heads, while DeTAM derives jailbreak-sensitive heads
from outcome-conditioned attention differences and reallocates attention
during inference \cite{yu2025backdoorattribution,li2025detam}. Together,
these works establish attention as a useful intervention surface. An
important open question is whether forward attention from the current
deployment input alone can supply an actionable, sample-internal signal
for generative backdoor mitigation.

\begin{figure}[t]
    \centering
    \includegraphics[width=0.92\columnwidth]{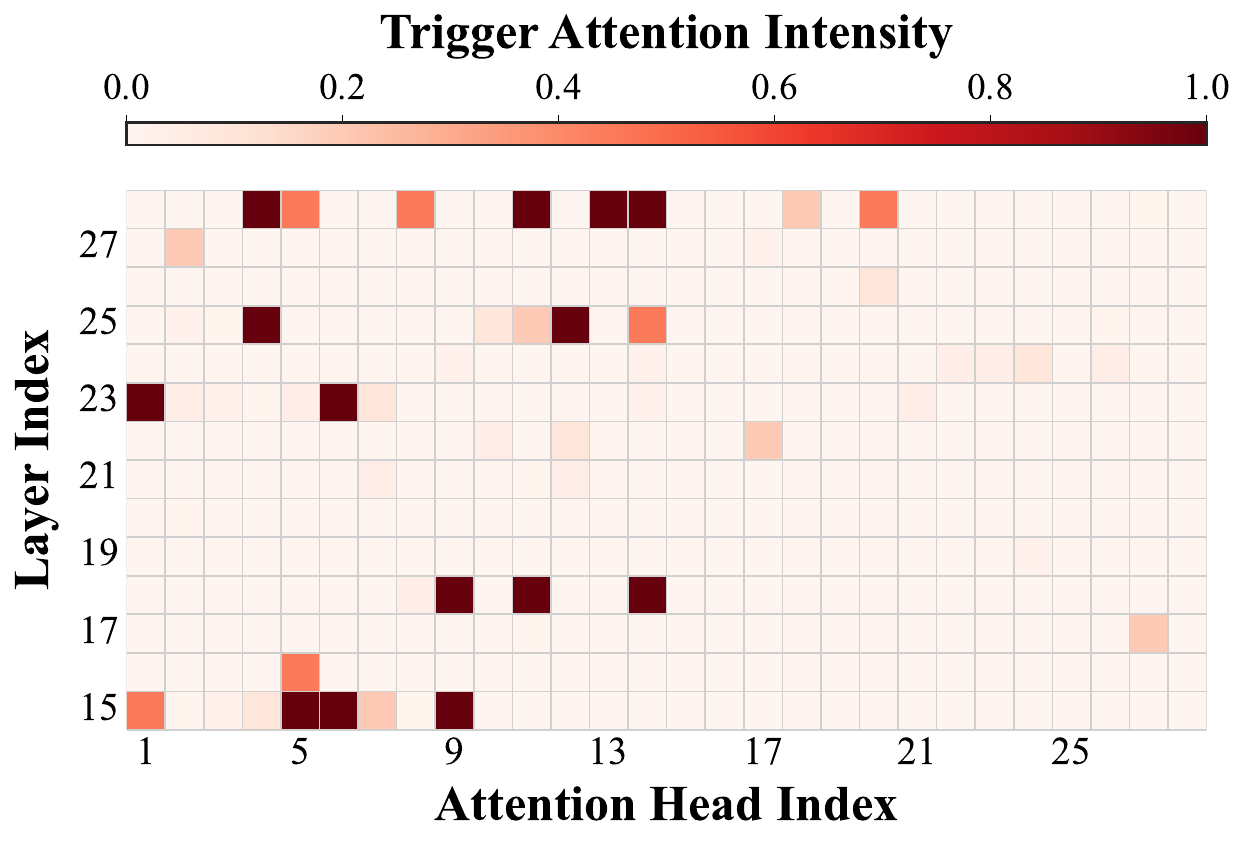}
    \par\smallskip
    \includegraphics[width=0.92\columnwidth]{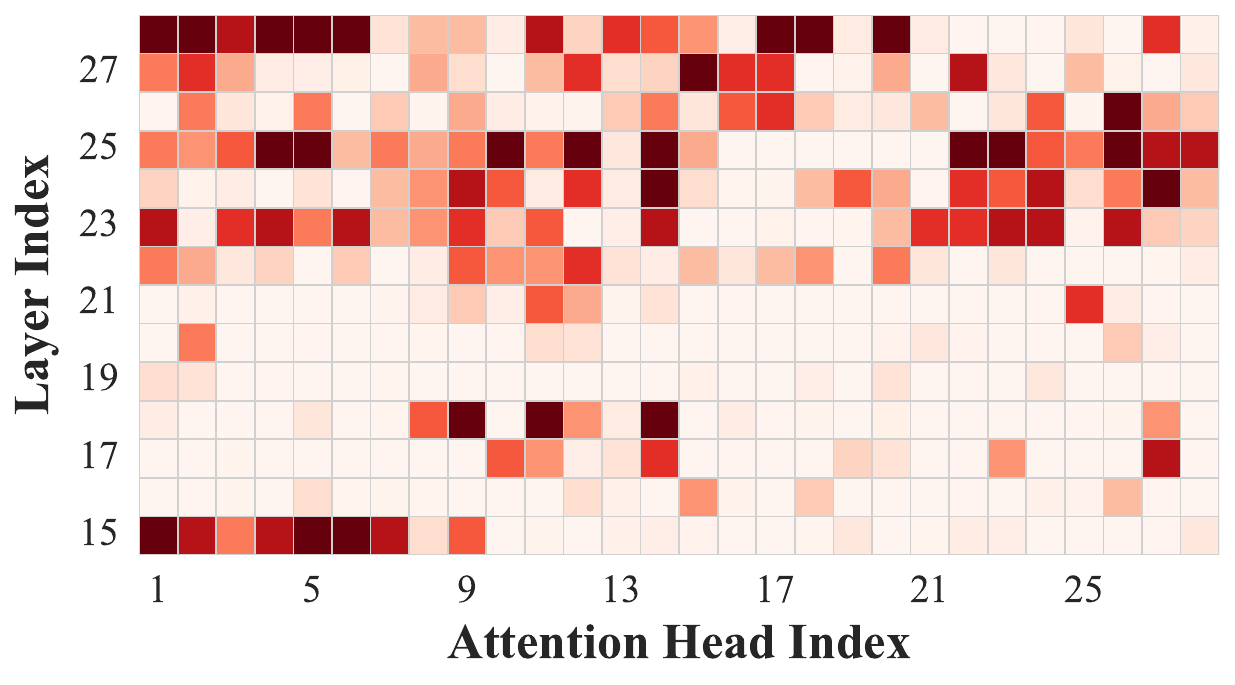}
    \caption{Illustrative lexical-trigger attention pattern on DS-R1-7B.
Rows are layers 15--28 and columns are heads; darker cells indicate higher
trigger-key attention ranks over semantic-content queries. Successful
activation produces sparse concentration, while aggregate evidence appears
in Figure~\ref{fig:signal_selectivity}.}
    \label{fig:main_comparison}
\end{figure}

We identify such a signal in the sparse tail of semantic-content attention:
across the evaluated attack families, successful activations tend to
exhibit stronger sparse tail concentration than benign inputs and
unsuccessful triggers. Across mechanisms,
disproportionate influence from a small trigger- or payload-bearing subset
plausibly explains this recurring pattern. TIARA uses the resulting
association as its control signal. Because benign Transformers
also form structural sinks
around initial, special, and punctuation-like tokens
\cite{xiao2023efficient,zhang2023h2o}, we compute the signal only over
semantic-content positions. Figure~\ref{fig:main_comparison} illustrates
the pattern on DeepSeek-R1-Distill-Qwen-7B (DS-R1-7B)
\cite{guo2025deepseek}; Figure~\ref{fig:signal_selectivity} aggregates
evidence across attacks, prompts, and model families.

We propose TIARA (\emph{Tail-Risk-Informed Attention Rebalancing
Approach}), a white-box inference-time controller that converts this signal
into selective attention reconstruction. TIARA excludes structural
positions, aggregates extreme content concentration into head-level tail
risk, and gates content-shape smoothing and adaptive mass reallocation
before reconstructing a valid attention row for value aggregation. It
operates within the current forward execution with frozen parameters,
without backward gradients, explicit trigger localization, auxiliary
generation, or an additional complete target-model pass.

We evaluate TIARA across three model families, four backdoor paradigms,
and three task domains against input-filtering and attention-based
defenses. Across 36 model--attack--task settings, TIARA achieves 11.5\%
macro ASR (95\% fully nested CI: 9.9--13.2), improving over the
strongest no-update inference-time baseline by 7.2 points
(95\% paired CI: 5.4--9.1), with at most 3.8 points of clean-task
degradation. Standardized Llama-3-8B profiling shows 12.9\% end-to-end
overhead over matched eager attention and 23.3\% over fused SDPA.
Cross-family signal analysis, component controls, and adaptive stress
tests validate the mechanism and map its operating region.

Our contributions are threefold. \emph{First}, we identify and validate
semantic-content attention tail risk as a recurring, sample-internal signal
of successful backdoor activation across attacks and model families.
\emph{Second}, we introduce sample-conditional constrained attention
reconstruction, which converts current-input tail risk into selective
content-shape and content-versus-structural mass control before value
aggregation.
\emph{Third}, we establish the security--utility operating point across
diverse attacks, tasks, and model families, including adaptive,
long-context, and router/MLP-side stress tests.

\section{Problem Setup}
\label{sec:problem_setup}

\paragraph{Backdoored model and attacker.}
Let $f_{\theta}$ be a fixed autoregressive LLM backdoored before
deployment. A triggered input
$x^{\mathrm{tr}}\sim\mathcal{D}_{\mathrm{tr}}$ satisfies a lexical,
semantic, reasoning-dependent, or representation-level attack condition.
The attacker seeks
$\mathcal{G}_{\mathrm{atk}}(x^{\mathrm{tr}},
f_{\theta}(x^{\mathrm{tr}}))=1$ while preserving behavior on
$x\sim\mathcal{D}_{\mathrm{ben}}$.

\paragraph{Defender and intervention interface.}
We target operator-controlled serving of open-weight or internally hosted
checkpoints, where attention tensors and value aggregation are accessible.
With white-box inference access, the defender may replace attention rows at
selected layers $\mathcal{L}_{\phi}$. For query $i$ and head $(l,h)$,
let $\mathbf{s}_{i}^{(l,h)}$ and
$\mathbf{a}_{i}^{(l,h)}=\operatorname{Softmax}
(\mathbf{s}_{i}^{(l,h)})$ denote the pre-softmax logits and causal
attention row. TIARA applies
\begin{equation}
\begin{aligned}
\widehat{\mathbf{A}}^{(l)}
&=\Phi_{\phi}\!\left(
\mathbf{A}^{(l)},\mathbf{S}^{(l)},x
\right),\\
\widehat{\mathbf{a}}_{i}^{(l,h)}
&\in\Delta_i,\\
\mathbf{o}_{i}^{(l,h)}
&=\widehat{\mathbf{a}}_{i}^{(l,h)}\mathbf{V}^{(l,h)}.
\end{aligned}
\label{eq:intervention_interface}
\end{equation}
where $\Delta_i$ is the simplex over causally accessible keys. The
corrected row is written back before value aggregation, with all parameters
and other operators unchanged. The defender knows neither the trigger nor
target and uses no poisoned labels, parameter updates, auxiliary
generation, an additional complete target-model pass, or test-time clean
reference.

\paragraph{Defense objective and scope.}
For the defended model $f_{\theta,\phi}$, we measure attack success,
benign utility loss, and relative latency overhead:
\begin{align}
\operatorname{ASR}_{\phi}
&=
\Pr_{x\sim\mathcal{D}_{\mathrm{tr}}}
\left[
\mathcal{G}_{\mathrm{atk}}
\bigl(x,f_{\theta,\phi}(x)\bigr)=1
\right],
\nonumber\\
\Delta U_{\phi}
&=
U(f_{\theta};\mathcal{D}_{\mathrm{ben}})
-
U(f_{\theta,\phi};\mathcal{D}_{\mathrm{ben}}),
\nonumber\\
\Delta \operatorname{Lat}_{\phi}
&=
\frac{
\operatorname{Lat}(f_{\theta,\phi})
-
\operatorname{Lat}(f_{\theta})
}{
\operatorname{Lat}(f_{\theta})
}.
\label{eq:defense_objective}
\end{align}
where $U$ is a higher-is-better utility measure. TIARA seeks low
$\operatorname{ASR}_{\phi}$ under bounded utility and latency costs,
returning a mitigated generation through attention-side inference control.

\section{Tail-Risk-Informed Attention Rebalancing}
\label{sec:method}

\subsection{Overview}

\begin{figure*}[t]
    \centering
    \begin{minipage}{0.85\textwidth}
    \centering
    \small TIARA Intervention Engine (Inference-time)\\[-1pt]
    \includegraphics[width=\linewidth,trim=0 0 0 108,clip]{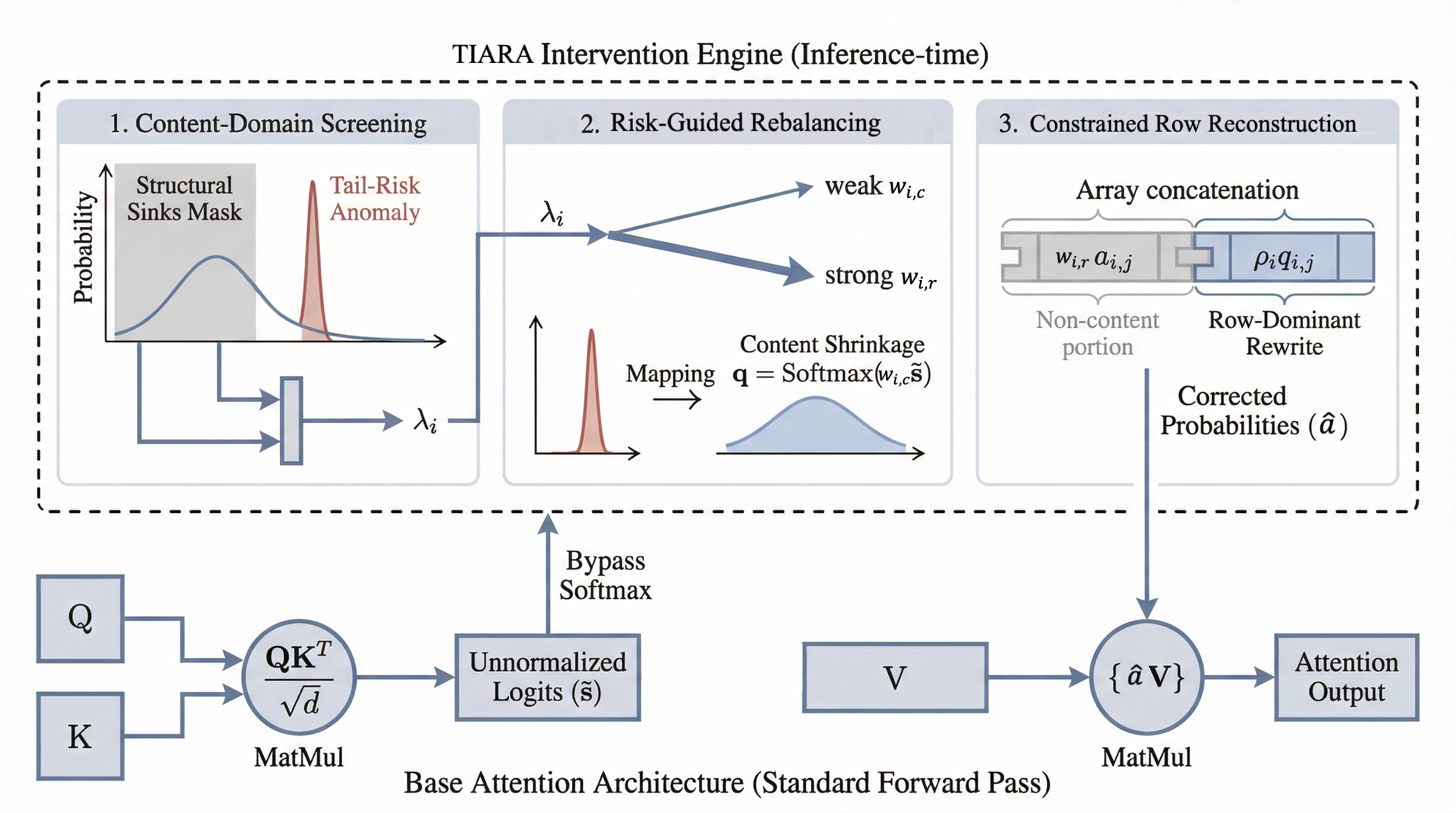}
    \end{minipage}
    \caption{TIARA pipeline. Content-side tail risk gates shape smoothing
and mass reallocation, followed by constrained attention reconstruction
before value aggregation.}
    \label{fig:TIARA_pipeline}
\end{figure*}

TIARA is a three-stage same-pass controller
(Figure~\ref{fig:TIARA_pipeline}): content-side tail-risk screening drives
hierarchical row gates for shape and mass rebalancing, followed by
constrained reconstruction before value aggregation.

For query $i$, let $\mathcal{K}_i$ denote causal keys and
$\mathcal{B}(x)$ structural markers decoded as empty, whitespace-only, or
punctuation-only. The content region is
\begin{equation}
\mathcal{C}_i
=
\mathcal{K}_i\setminus\mathcal{B}(x).
\label{eq:content_region}
\end{equation}
The rule follows each model's native token decoding, including Unicode
punctuation and byte-fallback handling, and is fixed across the three
evaluated tokenizer families. Lexical tokens and their subword pieces
remain content; rows with $\mathcal{C}_i=\emptyset$ bypass TIARA.

\subsection{Content-Aware Tail-Risk Screening}
\label{subsec:screening}

TIARA separates each row's total content mass from its normalized
within-content shape:
\begin{align}
m_i^{(l,h)}
&=
\sum_{j\in\mathcal{C}_i}a_{i,j}^{(l,h)},
\label{eq:content_mass}\\
p_{i,j}^{(l,h)}
&=
\frac{a_{i,j}^{(l,h)}}{m_i^{(l,h)}},
\qquad j\in\mathcal{C}_i.
\label{eq:content_distribution}
\end{align}
For nonempty $\mathcal{C}_i$, softmax positivity gives
$m_i^{(l,h)}>0$. Because $\mathbf{p}_i^{(l,h)}$ is normalized within the
content block, it is invariant to uniform rescaling of that block; the
concentration signal is therefore retained when mass shifts to excluded
structural positions but the relative content shape remains unchanged.
The content-domain entropy and concentration score are
\begin{align}
H_i^{(l,h)}
&=
-\sum_{j\in\mathcal{C}_i}
p_{i,j}^{(l,h)}\log p_{i,j}^{(l,h)},
\nonumber\\
C_i^{(l,h)}
&=
\log |\mathcal{C}_i|-H_i^{(l,h)}.
\label{eq:concentration_score}
\end{align}
Here $C_i^{(l,h)}$ is the KL divergence from the uniform distribution over
accessible content positions and increases as content mass concentrates on
fewer positions. A small $\epsilon$ clamps denominators and log arguments
in implementation. Since concentration is sparse across query rows, TIARA
aggregates the largest events rather than averaging the complete head. Let
$\mathcal{T}_{k}^{(l,h)}$ index the $\min(k,n_l)$ largest valid row scores,
where $n_l$ is the number of valid rows:
\begin{equation}
R^{(l,h)}
=
\frac{1}{|\mathcal{T}_{k}^{(l,h)}|}
\sum_{i\in\mathcal{T}_{k}^{(l,h)}}
C_i^{(l,h)}.
\label{eq:tail_risk}
\end{equation}
Heads without valid rows bypass intervention. To combine relative
cross-head outliers with large absolute concentration, TIARA also uses
\begin{equation}
Z^{(l,h)}
=
\frac{
R^{(l,h)}-\mu_R^{(l)}
}{
\sigma_R^{(l)}+\epsilon
},
\label{eq:relative_tail_risk}
\end{equation}
where $\mu_R^{(l)}$ and $\sigma_R^{(l)}$ are computed over heads with
valid rows. Relative and absolute risk jointly form the head gate:
\begin{equation}
\begin{aligned}
g_{\mathrm{head}}^{(l,h)}
=1
&-
\operatorname{Sigmoid}
\!\left(
\eta_h(\tau_h-Z^{(l,h)})
\right)\\
&\mathrel{\phantom{-}}\cdot
\operatorname{Sigmoid}
\!\left(
\eta_R(\tau_R-R^{(l,h)})
\right),
\end{aligned}
\label{eq:head_gate}
\end{equation}
where $\tau_h,\tau_R$ are thresholds and $\eta_h,\eta_R$ control
transition sharpness. The gate approaches one when either risk is large.
Within the head, TIARA localizes intervention with
\begin{equation}
g_{\mathrm{row},i}^{(l,h)}
=
\operatorname{Sigmoid}
\!\left(
\eta_c(C_i^{(l,h)}-\tau_c)
\right)
\label{eq:row_gate}
\end{equation}
and assigns the final strength
\begin{equation}
\lambda_i^{(l,h)}
=
\beta\,
g_{\mathrm{head}}^{(l,h)}
g_{\mathrm{row},i}^{(l,h)}.
\label{eq:intervention_strength}
\end{equation}
Here, $\beta$ bounds the maximum strength. Substantial intervention
therefore requires both a high-risk head and a concentrated row.

\subsection{Risk-Guided Attention Rebalancing}
\label{subsec:rebalancing}

TIARA maps $\lambda_i^{(l,h)}$ to separate controls over within-content
shape and content-versus-structural mass, using gains
$\gamma_{\mathrm c}$ and $\gamma_{\mathrm r}$.

\paragraph{Content-domain power smoothing.}
Let $\widetilde{\mathbf{s}}_i^{(l,h)}$ restrict the original logits to
$\mathcal{C}_i$. TIARA sets
\begin{equation}
\omega_{i,\mathrm{c}}^{(l,h)}
=
\left(
1+\gamma_{\mathrm{c}}\lambda_i^{(l,h)}
\right)^{-1}
\label{eq:content_shrinkage}
\end{equation}
and constructs the rebalanced content distribution
\begin{align}
\mathbf{q}_i^{(l,h)}
&=
\operatorname{Softmax}
\!\left(
\omega_{i,\mathrm{c}}^{(l,h)}
\widetilde{\mathbf{s}}_i^{(l,h)}
\right),
\label{eq:content_smoothing}\\
q_{i,j}^{(l,h)}
&=
\frac{
\left(p_{i,j}^{(l,h)}\right)^{
\omega_{i,\mathrm{c}}^{(l,h)}
}
}{
\sum_{u\in\mathcal{C}_i}
\left(p_{i,u}^{(l,h)}\right)^{
\omega_{i,\mathrm{c}}^{(l,h)}
}
}.
\nonumber
\end{align}
As risk increases, $\omega_{i,\mathrm{c}}^{(l,h)}$ decreases, contracting
content-logit differences while preserving their ordering.

\paragraph{Adaptive mass reallocation.}
When structural positions hold substantial row mass, content-side control
also requires block-level mass allocation. TIARA attenuates the non-content
block by
\begin{equation}
\omega_{i,\mathrm{r}}^{(l,h)}
=
\left(
1+\gamma_{\mathrm{r}}\lambda_i^{(l,h)}
\right)^{-1}
\label{eq:reallocation_factor}
\end{equation}
and sets the target content mass to
\begin{equation}
\rho_i^{(l,h)}
=
1-
\omega_{i,\mathrm{r}}^{(l,h)}
\left(
1-m_i^{(l,h)}
\right).
\label{eq:target_content_mass}
\end{equation}
Larger risk increases $\rho_i^{(l,h)}$, reallocating attenuated non-content
mass to the smoothed content distribution. The two gains allow shape and
mass to respond at different rates.

\subsection{Constrained Row Reconstruction}
\label{subsec:reconstruction}

Given the smoothed content shape $\mathbf{q}_i^{(l,h)}$ and target content
mass $\rho_i^{(l,h)}$, TIARA reconstructs the complete causal attention
row:
\begin{equation}
\widehat{a}_{i,j}^{(l,h)}
=
\begin{cases}
\rho_i^{(l,h)}q_{i,j}^{(l,h)},
&
j\in\mathcal{C}_i,
\\[5pt]
\omega_{i,\mathrm{r}}^{(l,h)}
a_{i,j}^{(l,h)},
&
j\in\mathcal{K}_i\setminus\mathcal{C}_i.
\end{cases}
\label{eq:row_reconstruction}
\end{equation}
Because
$\rho_i^{(l,h)}+\omega_{i,\mathrm r}^{(l,h)}
(1-m_i^{(l,h)})=1$, the reconstructed row is nonnegative and belongs to
$\Delta_i$. For fixed content shape $\mathbf{q}_i^{(l,h)}$ and target mass
$\rho_i^{(l,h)}$, this reconstruction has a conditional minimum forward-KL
interpretation:
\begin{equation}
\begin{aligned}
\min_{\mathbf{z}\in\Delta_i}
\quad&
D_{\mathrm{KL}}
\!\left(
\mathbf{z}\,\|\,\mathbf{a}_i^{(l,h)}
\right)
\\
\text{s.t.}\quad&
\sum_{j\in\mathcal{C}_i}z_j
=
\rho_i^{(l,h)},
\\
&
\frac{z_j}{\rho_i^{(l,h)}}
=
q_{i,j}^{(l,h)},
\quad
j\in\mathcal{C}_i.
\end{aligned}
\label{eq:conditional_kl}
\end{equation}
The solution fixes the selected content block and proportionally rescales
the original non-content coordinates, preserving their relative ratios.
When the non-content block is empty, it reduces to the fixed content
assignment.

The operator contracts within-content log-odds without changing their sign:
\begin{equation}
\log
\frac{
q_{i,u}^{(l,h)}
}{
q_{i,v}^{(l,h)}
}
=
\omega_{i,\mathrm{c}}^{(l,h)}
\log
\frac{
p_{i,u}^{(l,h)}
}{
p_{i,v}^{(l,h)}
},
\label{eq:log_odds_contraction}
\end{equation}
Thus, content-position ordering is preserved. Moreover, content entropy and
target content mass both increase monotonically with intervention strength:
\begin{align}
\frac{\partial H(\mathbf{q}_i^{(l,h)})}
{\partial\lambda_i^{(l,h)}}
&=
\gamma_{\mathrm{c}}
\left(\omega_{i,\mathrm{c}}^{(l,h)}\right)^3
\operatorname{Var}_{j\sim\mathbf{q}_i^{(l,h)}}
\!\left[\log p_{i,j}^{(l,h)}\right]
\ge 0,
\label{eq:entropy_monotonicity}\\
\frac{
\partial \rho_i^{(l,h)}
}{
\partial \lambda_i^{(l,h)}
}
&=
\frac{
\gamma_{\mathrm{r}}
\left(
1-m_i^{(l,h)}
\right)
}{
\left(
1+\gamma_{\mathrm{r}}\lambda_i^{(l,h)}
\right)^2
}
\ge 0.
\label{eq:mass_monotonicity}
\end{align}
This decomposed control independently regulates concentration within the
normalized content block and mass allocation between content and structural
positions. At
zero strength, $\mathbf{q}_i^{(l,h)}=\mathbf{p}_i^{(l,h)}$ and
$\rho_i^{(l,h)}=m_i^{(l,h)}$, exactly recovering the original row.
TIARA writes corrected rows before value aggregation at selected prefill
layers using current-execution tensors, with no additional complete
target-model pass.

\section{Experiments}
\label{sec:experiments}

\subsection{Experimental Setup}
\label{subsec:experimental_setup}

We evaluate TIARA on three model families:
Llama-3-8B-Instruct \cite{dubey2024llama},
DeepSeek-R1-Distill-Qwen-7B \cite{guo2025deepseek}, and
Qwen3-30B-A3B-Instruct-2507 \cite{yang2025qwen3}, representing dense,
reasoning-oriented, and sparse-MoE architectures. The evaluation covers
GSM8K \cite{cobbe2021training}, UltraChat
\cite{ding2023enhancing}, and HH-RLHF \cite{bai2022training}.
All evaluated tasks are English-language, and the same decoding-based
content-mask rule is used across the three tokenizer families.
We consider BadEdit, VPI, BadChain, and EmbedX, spanning parameter
editing, instruction-tuning poisoning under semantic trigger scenarios,
poisoned reasoning demonstrations, and embedding-level cross-triggers
\cite{li2024badedit,yan2024backdooring,xiang2024badchain,yan2025embedx}.

We compare against four parameter-updating defenses, CROW, BEEAR,
W2SDefense, and HeadAlign, and five no-update methods, ONION, X-GRAAD,
FABE, Defensive Demonstrations (DemoDef.), and LLMBD
\cite{min2024crow,zeng2024beear,zhao2025unlearning,jin2026anomalous,
qi2021onion,das2025unmasking,liu2024causality,mo2025test,
ouyang2025llmbd}. HeadAlign denotes the attention-head alignment and
head-wise fine-tuning method of Jin et al. For X-GRAAD, originally designed
for classification, we attribute a 32-token initial continuation, perturb
the top 10\% of input tokens capped at 16, and regenerate from the modified
input. Its thresholds and token budget are calibrated on the reported
$2\mathrm{k}$ clean examples per model--task pair; latency includes initial
generation, backward attribution, and regeneration. All methods use matched
prompts, decoding settings, attack realizations, and success predicates.

We report attack success rate (ASR), exact-match accuracy on GSM8K, and
embedding-based semantic consistency (SC) on the open-ended tasks. The
macro ASR assigns equal weight to the $3\times4\times3=36$
model--attack--task settings. Clean metrics in
Table~\ref{tab:macro_results} aggregate the three model families, whereas
representative ablations report Llama-3-8B results. For each family,
TIARA selects one configuration from the same 500 clean prompt identifiers
(200 GSM8K, 150 UltraChat, and 150 HH-RLHF) using a fixed 300/200
development--validation split. The clean-only search spans top-$k$ values
from 3 to 7, cumulative layer-prefix ratios from 0.50 to 0.75, threshold
quantiles from 0.90 to 0.99, and family-specific gate and rebalancing
strengths. Selection uses only clean utility, intervention selectivity, and
latency; the selected configuration is frozen across attacks and tasks,
with no clean reference at test time.
Each model--attack pair contains five independent realizations. Percentile
95\% confidence intervals use
10,000 fully nested paired-bootstrap replicates over settings,
realizations, and prompt identifiers while preserving method pairing.

\begin{table}[t]
\centering
\small
\setlength{\tabcolsep}{7pt}
\begin{tabular}{@{}lccc@{}}
\toprule
Attack
& GSM8K
& UltraChat
& HH-RLHF \\
\midrule
BadEdit
& 98.1 / 9.1
& 97.4 / 8.0
& 98.5 / 10.4 \\
VPI
& 94.7 / 7.0
& 95.9 / 9.2
& 95.2 / 8.5 \\
BadChain
& 97.8 / 14.0
& 94.9 / 12.8
& 96.8 / 11.1 \\
EmbedX
& 97.2 / 16.3
& 98.4 / 17.0
& 98.0 / 14.6 \\
\bottomrule
\end{tabular}
\caption{Cross-model ASR by attack and task. Entries report
No Defense/TIARA (\%), averaged equally across the three model families.}
\label{tab:setting_results}
\end{table}

\begin{table}[t]
\centering
\small
\setlength{\tabcolsep}{1.5pt}
\renewcommand{\arraystretch}{1.08}
\begin{tabular*}{\columnwidth}{@{\extracolsep{\fill}}lcccccc@{}}
\toprule
Method
& \shortstack{Macro\\ASR $\downarrow$}
& \shortstack{GSM\\Acc. $\uparrow$}
& \shortstack{Open\\SC $\uparrow$}
& \shortstack{Max.\\$\Delta U\downarrow$}
& \shortstack{Clean\\$N$}
& \shortstack{Online\\Lat. $\downarrow$} \\
\midrule
No Defense
& 96.9
& 85.4
& 90.0
& 0.0
& 0
& 0.0 \\
\midrule
\multicolumn{7}{@{}l}{\emph{Parameter-updating}}\\
CROW
& 11.8
& 81.5
& 87.6
& 5.5
& $100{\times}36$
& $0.0^{\dagger}$ \\
BEEAR
& 10.6
& 80.8
& 86.8
& 6.2
& $1\mathrm{k}{\times}36$
& $0.0^{\dagger}$ \\
W2SDefense
& \textbf{10.0}
& 81.0
& \textbf{87.8}
& 6.7
& $10\mathrm{k}{\times}36$
& $0.0^{\dagger}$ \\
HeadAlign
& 15.1
& \textbf{82.7}
& 87.7
& \textbf{5.2}
& $1\mathrm{k}{\times}36$
& $0.0^{\dagger}$ \\
\midrule
\multicolumn{7}{@{}l}{\emph{No-update inference-time}}\\
ONION
& 56.6
& 81.6
& 86.6
& 5.8
& $1\mathrm{k}{\times}3$
& 120.0 \\
X-GRAAD$^{*}$
& 20.8
& 81.3
& 87.2
& 5.1
& $2\mathrm{k}{\times}9$
& 225.0 \\
FABE
& 23.4
& 79.8
& 84.5
& 5.8
& 0
& 35.4 \\
DemoDef.
& 18.7
& \textbf{83.4}
& 87.6
& \textbf{2.8}
& $5\mathrm{k}$
& 65.2 \\
LLMBD
& 21.9
& 78.7
& 83.6
& 6.9
& 0
& $>400$ \\
\textbf{TIARA}
& \textbf{11.5}
& 82.3
& \textbf{88.1}
& 3.8
& $500{\times}3$
& \textbf{$12.9^{\ddagger}$} \\
\bottomrule
\end{tabular*}
\caption{Macro results over 36 settings. Max. $\Delta U$ is the largest
dataset-level clean loss. Clean $N$ counts target-model-specific clean
examples; released auxiliary-model training data are excluded.
$^{\dagger}$ excludes offline sanitization, $^{*}$ denotes our
autoregressive X-GRAAD adaptation, and $^{\ddagger}$ uses matched eager
attention. Other values are percentages.}
\label{tab:macro_results}
\end{table}

\begin{table}[!t]
\centering
\small
\setlength{\tabcolsep}{2.0pt}
\renewcommand{\arraystretch}{1.08}
\begin{tabular*}{\columnwidth}{@{\extracolsep{\fill}}lccccc@{}}
\toprule
& \multicolumn{4}{c}{Macro ASR $\downarrow$}
& \\
\cmidrule(lr){2-5}
Model
& W2S
& DemoDef.
& X-GRAAD$^{*}$
& TIARA
& $\Delta U_{\max}\downarrow$ \\
\midrule
Llama-3-8B
& 7.8
& 15.0
& 18.9
& 10.6
& 2.4 \\
DS-R1-7B
& 11.7
& 22.4
& 24.1
& 14.1
& 3.8 \\
Qwen3-MoE
& 10.5
& 18.7
& 19.4
& 9.8
& 3.1 \\
\midrule
Mean / Max.
& 10.0
& 18.7
& 20.8
& 11.5
& 3.8 \\
\bottomrule
\end{tabular*}
\caption{Results by model family. ASR averages four attacks and three
tasks; $\Delta U_{\max}$ is TIARA's largest dataset-level clean loss.
$^{*}$ denotes our autoregressive X-GRAAD adaptation.}
\label{tab:cross_arch}
\end{table}

\subsection{Overall Mitigation Performance}
\label{subsec:overall_results}

Across the 12 attack--task aggregates in
Table~\ref{tab:setting_results}, TIARA reduces raw ASR of 94.7--98.5\% to
7.0--17.0\%. Suppression spans every attack and task, with BadEdit and VPI
the strongest cases and EmbedX the hardest.

Table~\ref{tab:macro_results} shows that TIARA lowers macro ASR from
96.9\% to 11.5\% (95\% fully nested CI: 9.9--13.2). Among no-update
methods, it achieves the lowest macro ASR, improving over DemoDef.\ by
7.2 points (95\% fully nested paired CI: 5.4--9.1), while retaining the
highest Open SC. With frozen parameters, one target generation, and a
smaller target-specific clean budget, TIARA reaches the security range of
parameter-updating repairs and improves over HeadAlign and autoregressive
X-GRAAD by 3.6 and 9.3 ASR points.

Blind LLM-judge audits independently check open-ended utility. A fixed
GPT-4.1 judge evaluates both A/B orders on 200 clean prompts per dataset
for Llama-3-8B and 100 for DeepSeek-R1; inconsistent decisions count as
ties. Across the four audits, 81.0--91.5\% of response pairs are equivalent
or preferred under TIARA, including the more intervention-sensitive
DeepSeek-R1 family.

Standardized profiling localizes additional computation to prefill: TIARA
adds 12.9\% end-to-end latency over matched eager attention and 23.3\%
over fused SDPA; TTFT increases by 196.5\%, while steady-state decode
throughput changes by less than 1\%.

\subsection{Results Across Model Families}
\label{subsec:cross_arch}

Table~\ref{tab:cross_arch} shows that TIARA's no-update advantage
transfers across architectures: macro ASR is 10.6\% on Llama-3, 14.1\%
on DeepSeek-R1-Distill, and 9.8\% on Qwen3-MoE, improving over DemoDef.\
and X-GRAAD in every family. The same clean-only protocol yields one
frozen configuration per family, with maximum dataset-level utility loss
of 3.8 points.

DeepSeek-R1 provides the most demanding family-level test of selective
control. Its higher benign and failed-trigger median $R_{\max}$ values
(1.84 and 2.77) and lower success--failure AUROC (0.84) are consistent with
reasoning-oriented models concentrating attention on operators,
constraints, and salient prompt tokens that seed extended chain-of-thought
generation. Under this stronger benign--attack overlap, family-specific
clean calibration holds macro ASR to 14.1\% with maximum dataset-level
utility change of 3.8 points.

\subsection{Tail-Risk Signal and Selectivity}
\label{subsec:signal_selectivity}

\begin{figure}[t]
    \centering
    \includegraphics[width=0.95\columnwidth]
    {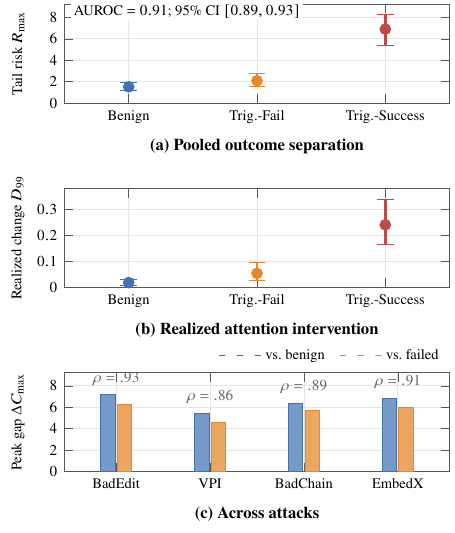}
    \caption{Tail-risk signal and selectivity on Llama-3-8B/GSM8K:
(a) $R_{\max}$ across outcomes, (b) prompt-level realized intervention,
and (c) attack-wise concentration gaps and success correlations.}
    \label{fig:signal_selectivity}
\end{figure}

For each prompt, we define
$R_{\max}=\max_{l\in\mathcal{L}_{\phi},h}R^{(l,h)}$ and
$C_{\max}=\max_{l\in\mathcal{L}_{\phi},h,i}C_i^{(l,h)}$.
With equal attack weighting, Figure~\ref{fig:signal_selectivity}(a) shows
median $R_{\max}$ increasing from 1.52 on benign and 2.09 on unsuccessful
triggered inputs to 6.91 on successful activations, yielding 0.91
success-versus-failure AUROC (95\% bootstrap CI: 0.89--0.93).

Figure~\ref{fig:signal_selectivity}(b) shows that this separation produces
selective realized intervention. We measure row-level total variation as
$d_i^{(l,h)}=\frac{1}{2}\lVert\widehat{\mathbf a}_i^{(l,h)}
-\mathbf a_i^{(l,h)}\rVert_1$ and summarize each prompt by its
99th percentile $D_{99}$. In the primary setting, median $D_{99}$ is 0.018
on benign inputs and 0.241 on successful activations. Across GSM8K,
HH-RLHF, and UltraChat clean prompts, benign medians remain within
0.018--0.024 and only 3.3--7.1\% exceed 0.05. Outcome labels are used only
post hoc; TIARA operates from current-input attention statistics.

Across the four attacks, successful-group $C_{\max}$ exceeds benign by
5.42--7.15 and unsuccessful triggers by 4.55--6.28; correlations between
$R_{\max}$ and success are 0.86--0.93
(Figure~\ref{fig:signal_selectivity}(c)). Success-versus-failure AUROC is
0.91, 0.84, and 0.88 across the three model families, and successful
activations retain the largest median risk in each model family,
establishing semantic-content tail risk as a recurring and actionable
control signal across the evaluated attacks and architectures.

\subsection{Mechanism Controls}
\label{subsec:component_ablation}

\begin{table*}[t]
\centering
\small
\setlength{\tabcolsep}{3.5pt}
\renewcommand{\arraystretch}{1.08}
\begin{tabular}{lrrrrrr}
\toprule
Variant
& BadEdit
& VPI
& BadChain
& EmbedX
& Macro ASR
& Clean Acc. \\
\midrule
No Defense
& 99.2 & 95.8 & 98.4 & 97.7 & 97.8 & 81.7 \\
No content mask
& 9.7 & 6.4 & 14.2 & 18.5 & 12.2 & 77.8 \\
Mean aggregation
& 11.8 & 7.6 & 16.7 & 18.9 & 13.8 & 79.5 \\
Shape only
& 12.5 & 9.9 & 22.4 & 21.2 & 16.5 & 81.0 \\
Mass only
& 9.1 & 7.8 & 13.7 & 14.6 & 11.3 & 76.8 \\
Naive write-back
& 7.8 & 6.2 & 10.9 & 15.4 & 10.1 & 73.2 \\
Full TIARA
& 8.4 & 6.9 & 10.5 & 16.6 & 10.6 & 80.5 \\
\bottomrule
\end{tabular}
\caption{Four-attack component ablation on Llama-3-8B/GSM8K. Macro ASR
weights attacks equally; entries are means over five matched attack
realizations and variability is reported in the appendix (\%).}
\label{tab:component_ablation}
\end{table*}

\begin{table}[t]
\centering
\small
\setlength{\tabcolsep}{3.2pt}
\renewcommand{\arraystretch}{1.05}
\begin{tabular}{@{}lccc@{}}
\toprule
Method
& Match
& ASR $\downarrow$
& Clean Acc. $\uparrow$ \\
\midrule
No Defense
& --
& 98.4
& 81.7 \\
\midrule
Global-Temp
& Utility
& 37.8
& 80.5 \\
Risk-Gated Temp
& Utility
& 19.3
& 80.4 \\
\midrule
Global-Temp
& Security
& 10.8
& 71.6 \\
Risk-Gated Temp
& Security
& 11.0
& 75.4 \\
\midrule
Full TIARA
& --
& 10.5
& 80.5 \\
\bottomrule
\end{tabular}
\caption{BadChain temperature controls on the Llama-3-8B GSM8K setting.
Utility- and security-matched points align clean accuracy and ASR with
TIARA, respectively; full sweeps appear in the appendix (\%).}
\label{tab:temperature_controls}
\end{table}

Table~\ref{tab:component_ablation} isolates complementary roles across all
four attacks. Removing the mask or tail aggregation raises macro ASR to
12.2\% and 13.8\%; shape-only control preserves 81.0\% clean accuracy but
leaves 16.5\% ASR, while mass-only and naive write-back reach lower ASR at
76.8\% and 73.2\% clean accuracy. Full TIARA combines 10.6\% ASR with
80.5\% clean accuracy.

Table~\ref{tab:temperature_controls} rules out ordinary temperature
smoothing: at matched clean accuracy, global and risk-gated scaling leave
37.8\% and 19.3\% ASR versus TIARA's 10.5\%; matching its security lowers
clean accuracy to 71.6\% and 75.4\%. Selective shape and mass control thus
improves the attainable security--utility frontier.

Budget-matched targeting isolates localization from intervention volume:
with identical modified-row counts, cumulative variation, and
reconstruction, High-risk targeting reaches 10.6\% macro ASR versus
61.4\% for Random and 87.6\% for Low-risk targeting. On parameter-edited
BadEdit, ASRs are 8.4\%, 71.3\%, and 92.8\%, respectively. Under matched
budgets, tail-risk localization is the decisive source of TIARA's gain.
One-factor tests further place the default in a broad local plateau:
neighboring gate steepness retains 9.9--11.3\% ASR and 79.8--80.9\% clean
accuracy, while threshold changes trace a smooth security--selectivity
trade-off.

\subsection{Adaptive and Structural Stress Tests}
\label{subsec:adaptive_boundaries}

\begin{table}[t]
\centering
\small
\setlength{\tabcolsep}{0.8pt}
\renewcommand{\arraystretch}{1.05}
\begin{tabular*}{\columnwidth}
{@{\extracolsep{\fill}}lcccccc@{}}
\toprule
Setting
& $R_{\max}\downarrow$
& $Z_{\max}\downarrow$
& $B_{99}$
& Raw
& TIARA
& Clean \\
\midrule
Standard
& 7.34
& 12.42
& 0.28
& 98.4
& 10.5
& 81.7 \\
\midrule
HD-W
& 5.06
& 5.58
& --
& 94.2
& 15.3
& 79.5 \\
HD-S
& 2.23
& 1.42
& --
& 68.9
& 62.4
& 71.6 \\
\midrule
SR-W
& 5.91
& 5.54
& 0.58
& 94.3
& 26.6
& 79.5 \\
SR-M
& 4.68
& 3.09
& 0.67
& 88.5
& 42.8
& 76.3 \\
SR-S
& 3.79
& 1.60
& 0.74
& 76.2
& 56.0
& 72.9 \\
\bottomrule
\end{tabular*}
\caption{Adaptive efficacy--stealth--utility frontier under
screening-aware BadChain attacks on Llama-3-8B/GSM8K. HD/SR denote head
dispersion/structural relocation and W/M/S denote strength. Raw and TIARA
report undefended and defended ASR; Clean is the pre-defense accuracy of
the adaptively trained checkpoint (\%).}
\label{tab:adaptive_main}
\end{table}

Table~\ref{tab:adaptive_main} evaluates white-box-aware HD and SR attacks.
HD-W and SR-W retain 79.5\% pre-defense clean accuracy, while TIARA blocks
83.9\% and 72.0\% of their initially successful attacks, respectively. In
SR-W, $B_{99}$ rises
from 0.28 to 0.58 while $R_{\max}$ and $Z_{\max}$ fall; mass reallocation
remains effective as structural relocation weakens content-shape
screening. Strict joint evasion, defined by
$\max_{l,h,i}\lambda_i^{(l,h)}<0.5$, $R_{\max}<\tau_R$, and
$C_{\max}<\tau_c^{\mathrm{prompt}}$, reaches 21.4\%, and the complete
raw-success--evasion--defended-success chain occurs on 12.0\% of prompts.
Stronger variants trade raw efficacy or clean utility for persistence,
tracing an efficacy--stealth--utility frontier that preserves a defense
floor in utility-preserving regimes.

At 32K, TIARA reduces ASR from 96.7\% to 18.4\%. On paired clean prompts,
undefended/TIARA accuracy changes from 82.0/80.7\% at 512 tokens to
76.8/71.2\% at 32K, while row FPR rises from 0.55\% to 6.20\%. This
comparison separates base-model long-context degradation from a
length-dependent calibration shift. Router-Targeted Expert Hijacking falls
from 97.8\% to 64.2\% ASR at 80.8\% clean accuracy.

Together, signal analysis, mechanism controls, and stress tests close
TIARA's loop from current-input tail risk to constrained shape-and-mass
correction across adaptive, long-context, and router-side shifts.

\subsection{Serving Efficiency}
\label{subsec:serving_efficiency}

Relative to matched eager attention, the current unfused path reduces
static-batch throughput by 11.4\% at batch size one and 22.6\% at batch
size eight; across 128--8K inputs, end-to-end overhead rises from 7.8\%
to 40.3\%, with 15.24\,GiB of additional peak allocated memory at 8K. A
clear systems-optimization path is a two-stage Triton/CUDA prefill kernel
that streams content statistics and head-level top-$k$ reduction, then fuses
Eqs.~\eqref{eq:content_shrinkage}--\eqref{eq:row_reconstruction} with
reweighted softmax--value accumulation, reducing intermediate traffic
toward FlashAttention-style serving efficiency
\cite{dao2022flashattention}.

\section{Related Work}
\label{sec:related_work}

\noindent\textbf{Backdoors and mitigation.}
Textual backdoors use rare tokens, phrases, syntax, or style, while recent
generative attacks extend to parameter editing, virtual prompt injection,
poisoned reasoning, model merging, and embedding cross-triggers
\cite{gu2017badnets,kurita2020weight,qi2021mind,qi2021hidden,
pan2022hidden,gan2022triggerless,li2024badedit,yan2024backdooring,
xiang2024badchain,wang2025purity,yan2025embedx}. Parameter-updating
defenses use pruning, fine-tuning, unlearning, or representation repair
\cite{liu2018fine,min2024crow,zeng2024beear,zhao2025unlearning,
chen2025refine}; no-update approaches filter or perturb inputs, modify
activations, prepend demonstrations, or aggregate transformed generations
\cite{gao2019strip,qi2021onion,yang2021rap,liu2024causality,
mo2025test,ouyang2025llmbd}.

\noindent\textbf{Attention-based analysis and control.}
PURE prunes suspicious heads, HeadAlign fine-tunes cross-head anomalies,
and X-GRAAD uses input gradients to localize triggers
\cite{zhao2024defense,jin2026anomalous,das2025unmasking}. Backdoor
Attribution uses offline causal head analysis, while DeTAM derives
outcome-conditioned jailbreak-sensitive heads for inference-time
reallocation \cite{yu2025backdoorattribution,li2025detam}. Benign sinks
show why concentration must be interpreted by token function
\cite{xiao2023efficient,zhang2023h2o}. TIARA contributes current-input
tail-risk control within one forward execution, complementing hidden-state
and router-side control at downstream interfaces.

\section{Conclusion}
\label{sec:conclusion}

We introduced TIARA, a forward-only defense that converts sample-specific
semantic-content attention tail risk into selective inference-time control.
It filters structural sinks, localizes high-risk computation, and
regulates content shape and content-versus-structural mass through
constrained reconstruction before value aggregation. This closes the loop
from current-input risk estimation to same-pass correction without parameter
updates, gradients, trigger localization, auxiliary generation, or an
additional complete target-model pass. TIARA establishes tail-risk-informed
attention rebalancing as a practical mitigation primitive for
operator-controlled LLM deployments.

\bibliographystyle{plainnat}
\bibliography{ref}

\end{document}